\documentclass[aps,prl,reprint,letterpaper,showpacs,twocolumn,nofootinbib,floatfix,superscriptaddress]{revtex4-1}
\usepackage[UKenglish]{babel}
\usepackage{graphicx}
\usepackage[pdftex                     
           ,pagebackref=false          
           ,colorlinks=false           
           ]{hyperref}
\hypersetup{linkcolor= blue,           
            citecolor= red,            
            urlcolor=black}            

\usepackage{braket}
\usepackage{amsmath}
\usepackage{amssymb}
\usepackage{amsfonts}
\usepackage{mathtools}
\usepackage{siunitx}

\usepackage{times}

\usepackage{color}
\usepackage{capt-of}

\newcommand*{\Beff}{\vec{B}_\mathrm{eff}}
\newcommand*{\beff}{B_\mathrm{eff}}
\newcommand*{\Beffz}{B_\mathrm{eff}^z}

\newcommand{\xc}{\Delta_{\mathrm{xc}}}
\newcommand{\soi}{\alpha_{\mathrm{R}}}
\newcommand{\effmass}{m_{\mathrm{eff}}^\ast}
\newcommand{\emass}{m_{\mathrm{e}}}

\newcommand{\effmagbohr}{\mu_{\mathrm{B}}^\ast}

\newcommand{\com}{M_{\mathrm{com}}}
\newcommand{\tom}{M_{\mathrm{tom}}}
\newcommand{\itom}{m_{\mathrm{tom}}}
\newcommand{\icom}{m_{\mathrm{com}}}
\newcommand{\Om}{\mathbf{M}_\mathrm{om}}
\newcommand{\Com}{\mathbf{M}_{\mathrm{com}}}
\newcommand{\Tom}{\mathbf{M}_{\mathrm{tom}}}

\newcommand{\oms}{\chi_\mathrm{oms}}
\newcommand{\chiLP}{\chi_\mathrm{LP}^{\uparrow + \downarrow}}

\newcommand{\Gret}{G^{\mathrm{R}}}
\newcommand{\bsigma}{\boldsymbol{\sigma}}
\newcommand{\dd}{\mathrm{d}}
\newcommand{\tr}{\mathrm{tr}}
\newcommand{\id}{\mathrm{id}}
\renewcommand{\vec}[1]{\mathbf{#1}}
\newcommand{\hatn}{{\hat{\vec n}}}
\newcommand{\sgn}{\mathrm{sgn}}

\setcitestyle{super,open={},close={}}

\makeatletter
\renewcommand\@biblabel[1]{#1.}
\makeatother

\begin{document}


	\begin{abstract}
		{\bf 
			
			Electrons which are slowly moving through chiral magnetic textures can effectively be described as if they where influenced by electromagnetic fields emerging from the real-space topology. This adiabatic viewpoint has been very successful in predicting physical properties of chiral magnets. Here, based on a rigorous quantum-mechanical approach, we unravel the emergence of chiral and topological orbital magnetism in one- and two-dimensional spin systems. We uncover that the quantized orbital magnetism in the adiabatic limit can be understood as a Landau-Peierls response to the emergent magnetic field. Our central result is that the spin-orbit interaction in interfacial skyrmions and domain walls can be used to tune the orbital magnetism over orders of magnitude by merging the real-space topology with the topology in reciprocal space.
			Our findings point out the route to experimental engineering of orbital properties of chiral spin systems, thereby paving the way to the field of chiral orbitronics.  
		}
	\end{abstract}

\setcounter{secnumdepth}{3}
 \title{Engineering Chiral and Topological Orbital Magnetism of Domain Walls and Skyrmions}
 
 \author{Fabian R. Lux}
  \email{f.lux@fz-juelich.de}
  \address{Peter Gr\"unberg Institut and Institute for Advanced Simulation,\\Forschungszentrum J\"ulich and JARA, 52425 J\"ulich, Germany}
   \address{RWTH Aachen University, 52062 Aachen, Germany}
 \author{Frank Freimuth}
 \author{Stefan Bl\"ugel}
    \address{Peter Gr\"unberg Institut and Institute for Advanced Simulation,\\Forschungszentrum J\"ulich and JARA, 52425 J\"ulich, Germany}
 \author{Yuriy Mokrousov}
   \address{Peter Gr\"unberg Institut and Institute for Advanced Simulation,\\Forschungszentrum J\"ulich and JARA, 52425 J\"ulich, Germany}

 \address{Institute of Physics Johannes Gutenberg-University Mainz, 55128 Mainz, Germany}

 \maketitle


%
%

\null\cleardoublepage


The field of magnetism is witnessing a recent spark of interest in Berry phase and transport effects which originate in non-collinear magnetism and spin chirality~\cite{Nagaosa2013,Suergers2014,Nayak2016}. One of the recent outstanding observations made in this field is the generation of large current-induced Hall effects in strongly frustrated metallic antiferromagnets~\cite{Zhang2016} and the topological Hall effect (THE) in skyrmions~\cite{Neubauer2009,Suergers2014}. On the other hand,
the physics of the fundamental phenomenon of orbital magnetism has been experiencing a true revival which can be attributed to the advent of Berry phase concepts in condensed matter~\cite{Xiao2005,Xiao2010}. The Berry phase origin of the orbital magnetization (OM) and its close relation to the Hall effect makes us believe that  non-collinear spin systems can reveal a rich landscape of orbital magnetism relying on spin chirality rather than spin-orbit interaction (SOI)~\cite{Hoffmann2015,Dias2016,Hanke2016,Hanke2017}. The corresponding phenomenon of {\it topological} orbital magnetization (TOM)\cite{Hoffmann2015,Dias2016,Hanke2016,Hanke2017} is rooted in the same physical mechanism that drives the emergence of non-trivial transport properties such as the THE in chiral skyrmions or the anomalous Hall effect in chiral antiferromagnets ~\cite{Shindou2001,Chen2014,Kubler2014}. 

The promises of topological contribution to the orbital magnetization are seemingly very high, since it offers new prospects in influencing and detecting the chirality of the underlying spin texture by addressing the orbital degree of freedom, which is the central paradigm in the advancing field of {\it orbitronics}~\cite{Go2017}. 
And while the emergence of topological orbital magnetism in several $\si{\nano\meter}$-scale chiral systems has been shown from first principles and tight-binding calculations~\cite{Hoffmann2015,Dias2016}, our understanding of this novel phenomenon is basically absent. In particular, this concerns its conceptually clear definition as well as our ability to tailor the properties of this effect in complex interfacial chiral systems, which often exhibit strong spin-orbit interaction. These are the two central questions we address in this work.

As has been shown in the case of skyrmions, the variety of topological phenomena which arise intrinsically from the non-trivial magnetization configuration $\hatn(x,y)$ can be attributed to an ``emergent" magnetic field $\Beffz$~\cite{Nagaosa2013}. The occurrence of this field is connected to the gauge-invariant Berry phase the electron's wavefunction acquires when traversing the texture~\cite{Schulz2012,Freimuth2013,Everschor-Sitte2014} (see Fig.~\ref{figure1:emergent_fields}b) for an intuitive illustration). In the adiabatic limit, this phase can be attributed to the effect of $\Beffz$, explicitly given by the expression
\begin{equation}
    \Beffz = \pm \frac{\hbar}{2	e}\ \hatn \cdot \left( \frac{\partial \hatn}{\partial x} \times \frac{\partial \hatn}{\partial y} \right),
\label{eq:Beff}
\end{equation}
where the sign depends on the spin of the electron.
When integrated over an isolated skyrmion, the total flux of $\Beffz$ is quantized to integer multiples of $2 \Phi_0$, where
$\Phi_0 \approx \SI{2e+3}{\tesla\nano\meter\squared}$
is the magnetic flux quantum, while the integer prefactor can be identified with the topological charge of a skyrmion,
$N_\mathrm{sk}$, essentially counting the number of times the spin evolves around the unit sphere when traced along a path enclosing the skyrmion center.

Formally, the non-collinear system $\hatn(x,y)$ can therefore be portrayed as a collinear one, albeit at the price of introducing the magnetic field $\Beffz$ into the Schr\"odinger equation. Just as an ordinary magnetic field would, the emergent magnetic field in chiral systems couples directly to the orbital degree of freedom
and provides an intuitive mechanism for the topological Hall effect of skyrmions~\cite{Bruno2004} as well as a possible explanation for the emergence of TOM. \vfill 
{
    \onecolumngrid
    \vspace{6mm}
    \includegraphics[width=2.0\linewidth]{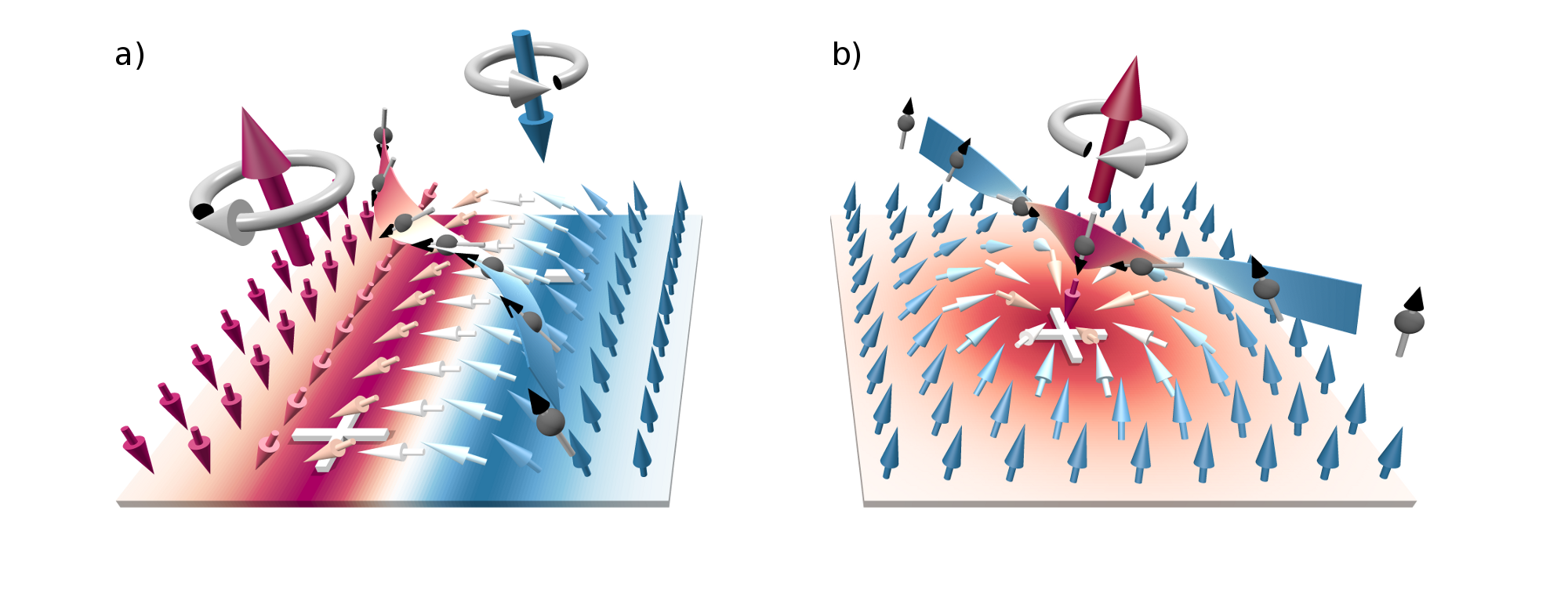}
    \captionof{figure}{Schematic depiction of emergent magnetic fields in a) N\'eel spirals and b) N\'eel skyrmions. As electrons (grey spheres) are adiabatically traversing these non-collinear magnetic textures (small arrows, with color indicating the $z$-projection), their wave function twists just in the same way as it would under the influence of an external magnetic field (the direction is depicted with vertical arrows, the sign and magnitude is illustrated by the colored background). The integrated flux of this emergent topological field over the skyrmion is quantized, while the averaged value of the emergent chiral field for a uniform spin-spiral is zero (although it can be non-zero for a 90$^{\circ}$~domain wall). The emergent field locally gives rise to persistent currents (depicted with circular arrows) and the corresponding a) chiral (for a spiral) and b) topological (for a skyrmion) orbital magnetization.}\label{figure1:emergent_fields}
    \twocolumngrid
}
\null\clearpage

Here, we uncover the emergence of distinct contributions to the orbital magnetization in slowly-varying chiral textures by following the intuition that such contributions should acquire the natural form 
\begin{equation}
    \vec{M}_{\rm orbital} \propto \oms~\Beff,
\label{eq:tom_hypothesis}
\end{equation}
where $\oms$ is the orbital magnetic susceptibility of the electronic system~\cite{Fukuyama1971,OgataMasao2015}. Indeed, we demonstrate that in the limit of vanishing SOI the topological orbital magnetization can be expressed in this way. We also discover that in the limit of small, yet non-zero SOI there is a novel {\it chiral} contribution to the orbital magnetization described by (\ref{eq:tom_hypothesis}) with the properly defined {\it chiral} emergent field which can be finite already for one-dimensional systems (see Fig.~\ref{figure1:emergent_fields}a)).

Moreover, by exploiting a rigorous semiclassical framework, we demonstrate that in interfacial chiral systems with finite SOI, the orbital magnetism can be tuned over orders of magnitude by varying the SOI strength within the range of experimentally observed values. We also underpin the crucial role that the topology of the local electronic structure of textures has in shaping the properties of orbital magnetism in chiral magnets. We discuss the bright avenues that our findings open, paving the way to the experimental observation of this phenomenon and to the exploitation of the orbital degree of freedom in chiral systems for the purposes of {\it chiral orbitronics}.



\begin{figure*}
 \centering
 \includegraphics[width=\linewidth]{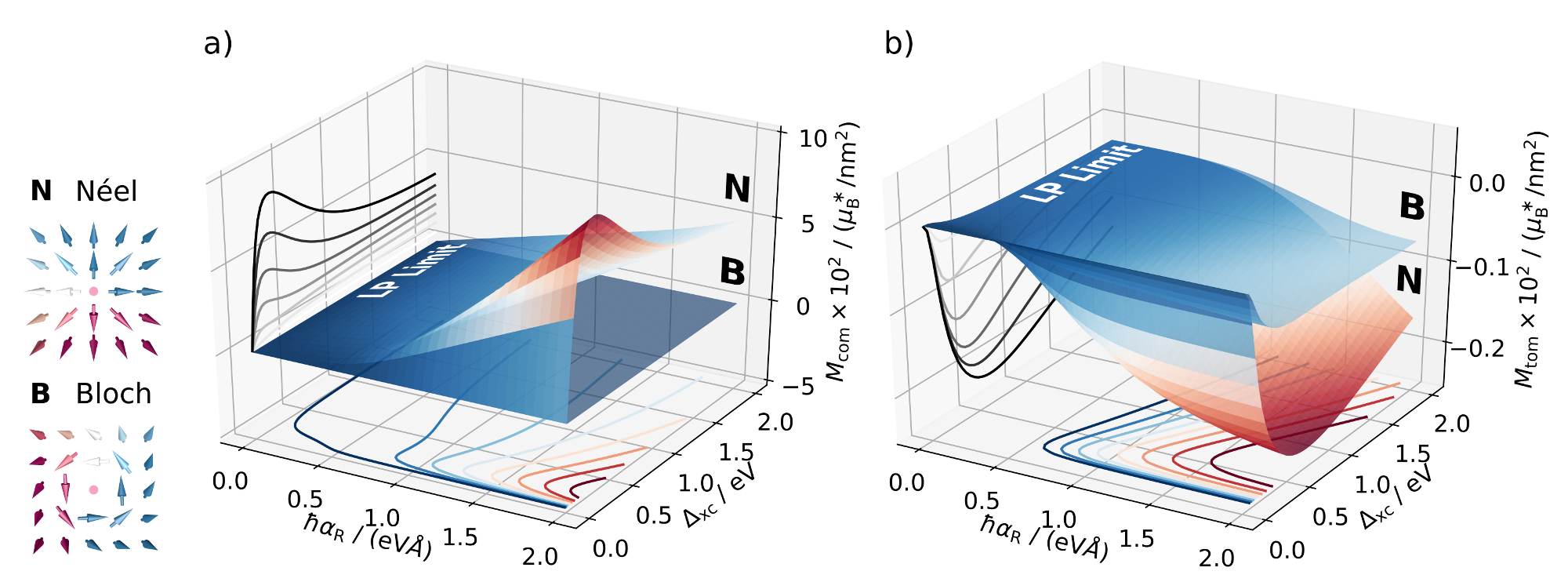}
 \caption{The phase diagram of a) chiral and b) topological orbital magnetization $\tom$, Eq.~(\ref{eq:com}) and Eq.~(\ref{eq:tom}), evaluated at the core of a N\'eel (Bloch) skyrmion ($m=1$, $c=0$\,nm, $w=20$\,nm) as a function of the parameters $\xc$ and $\soi$ of the Rashba Hamiltonian Eq.~(\ref{rashba:ham}) with $\mu=0$. The limit 
 $\xc \gg \soi$, $\xc \gtrsim 0.5\SI{1}{\electronvolt}$ corresponds to the coupling of the emergent magnetic field to the diamagnetic Landau-Peierls susceptibilty (what we refer to as ``LP limit"). 
 In an intermediate regime of $\xc \lesssim \soi$ orbital magnetism is strongly enhanced. 
 }
 \label{fig:phasediag}
\end{figure*}

\vspace{0.4cm}
\noindent{\bf Results}\\
The semiclassical formalism we are referring to in our work is based on the Green's function perturbation theory as presented by \citeauthor{Onoda2006}\cite{Onoda2006}. We put the orbital magnetism of chiral systems on a firm quantum-mechanical ground, formulating a rigorous theory for the emergence of orbital magnetism in non-collinear systems. The motivation for this approach is twofold. First of all, the expression for $\Beff$ arises from the adiabatic limit~\cite{Bruno2004,Fujita2011}, a regime where semiclassical approaches have been successfully applied in order to investigate Berry phase physics~\cite{Xiao2010}. Secondly, this certain type of gradient expansion~\cite{Rammer1986} provides a systematic guide through higher orders of perturbation theory where standard methods would be cumbersome.

It is based on an approximation to the single-particle Green's function and allows us to trace the orders of perturbation theory for chiral magnetic textures, distinguishing corrections to the out-of-plane orbital magnetization~\cite{Zhu2012} $\Om  = \hbar^1 M (\hat{\mathbf{n}}) \mathbf{e}_z$ of a locally ferromagnetic system which appear as powers of the derivatives of the magnetization with respect to real-space coordinates:
\begin{align}
 & \Com  =  \hbar^2 M^\alpha_i (\hat{\mathbf{n}})( \partial_i  n_\alpha )\mathbf{e}_z
 \label{eq:com}
\\
& \Tom  =  \hbar^3 M^{\alpha\beta}_{ij} (\hat{\mathbf{n}})( \partial_i  n_\alpha )( \partial_j  n_\beta )\mathbf{e}_z,
\label{eq:tom}
\end{align}
where $\partial_i = \partial / \partial x^i$. Here and in the following discussion, summation over repeated indices is implied with greek indices $\alpha,\beta \in \lbrace x,y,z \rbrace $ and latin indices $ i,j \in \lbrace x,y \rbrace$. 

The assignment of $\Tom$ to the second order expansion, Eq.~(\ref{eq:tom}), is based on our intuitive expectation, Eq.~(\ref{eq:tom_hypothesis}). The question whether or not this term is ``topological" will be discussed in the following and is answered by the semiclassical perturbation theory (see Methods). In addition to the effect of TOM we propose a novel contribution to the orbital magnetization, which is linear in the derivatives of the underlying texture, Eq.~(\ref{eq:com}), and thereby generally sensitive to its chirality. We thus refer to it as the \itshape chiral orbital magnetization \normalfont (COM). We will show how this effect can be attributed to a different kind of effective field (see Fig.~\ref{figure1:emergent_fields}a)) which emerges from the interplay of spin-orbit coupling and non-collinearity along one spatial dimension. 

While our approach is very general, for the purposes of including into consideration the effect of interfacial spin-orbit coupling and providing realistic numerical estimates, we focus our further analysis on the two-dimensional magnetic Rashba model 
\begin{equation}\label{rashba:ham}
H = \frac{\vec{p}^2}{2 \effmass} + \soi ( \boldsymbol{\sigma} \times \vec{p}  )_z + \xc\ \boldsymbol{\sigma} \cdot \hat{\mathbf{n}}(\vec{x}),
\end{equation}
where $\effmass$ is the electron's (effective) mass, $\boldsymbol{\sigma}$ denotes the vector of Pauli matrices, $\soi$ is the Rashba spin-orbit coupling constant, and $\xc$ is the strength of the local exchange field. This model has been proven to be extremely fruitful in unravelling various phenomena in surface magnetism~\cite{Manchon2015} and is known for its pronounced orbital response~\cite{Schober2012}.


\vspace{0.4cm}
\noindent{\bf Emergent Fields of Spin  Textures}.
Before discussing the emergence of orbital magnetism in this model, it is rewarding to discuss the appearance of effective fields in slowly-varying chiral spin textures in the limit of $|\soi| \ll |\xc|$. In this regime, it can be shown that to linear order in $\soi$, the spin-orbit coupling can be absorbed into a perturbative correction of the canonical momentum
$\vec{p} \to \vec{p} + e \boldsymbol{\mathcal{A}}^R$, with 
$\boldsymbol{\mathcal{A}}^R \equiv  \effmass \soi \epsilon^{ijz}  \sigma_i \vec{e}_j/ e$. This means that the Hamiltonian can be rewritten as:
\begin{equation}
H = \frac{ ( \vec{p} + e \boldsymbol{\mathcal{A}}^R) ^2 } {2 \effmass} + \xc \hatn \cdot \bsigma + \mathcal{O}(\soi^2). 
\end{equation}
For $|\soi| \ll |\xc|$\footnote{
To be precise, with correct physical dimensions, one should compare the length scales $\lambda_\mathrm{R} = \hbar / {\soi \effmass}$ and $\lambda_\mathrm{xc}=\hbar/{\sqrt{\xc \effmass}}$ and write $\lambda_\mathrm{R}\gg\lambda_\mathrm{xc}$ instead.
} and $\xc\to\infty$ the spin polarization of the wavefunctions is only weakly altered away from $\hatn$ and we can use an $SU(2)$ gauge field, defined by $\mathcal{U}^\dagger  (\boldsymbol{\sigma} \cdot \hat{\mathbf{n}})\, \mathcal{U} \equiv \sigma_z$, to rotate our Hamiltonian into the local axis specified by $\hatn$ (neglecting the terms of the order $\mathcal{O}(\soi^2)$)~\cite{Kim2013,Nakabayashi2014}: 
\begin{equation}
H \to \mathcal{U}^\dagger H \mathcal{U}= \frac{ \left(
\vec{p} + e  \boldsymbol{\mathcal{A}}(\mathbf{X})
\right)^2
 }{2 \effmass} +  \xc\ \sigma_z,
\end{equation}
where the potential $\boldsymbol{\mathcal{A}}$ now comprises the mixing of two gauge fields:
$\boldsymbol{\mathcal{A}} = \mathcal{U}^\dagger \boldsymbol{\mathcal{A}}^\mathrm{R} \,\mathcal{U}+ \boldsymbol{\mathcal{A}}^\mathrm{ xc}$, with the additional contribution $\boldsymbol{\mathcal{A}}^\mathrm{ xc} =- i\hbar \,\mathcal{U}^\dagger \nabla \mathcal{U} / e$. 
The essential idea is now the following: as $\xc\to\infty$, electrons are confined to the bands which correspond either to spin-up states $\ket{\uparrow}$ or spin-down states $\ket{\downarrow}$ depending on $\sgn(\xc)$.
This means that we can effectively replace the vector potential by its adiabatic counterpart, i.e.,
\begin{align}
\boldsymbol{\mathcal{A}} \rightarrow
\boldsymbol{\mathcal{A}}_\mathrm{ad} &\equiv
\sgn(\xc) \bra{\downarrow} \boldsymbol{\mathcal{A}} \ket{\downarrow}
\notag \\
&= \boldsymbol{\mathcal{A}}_\mathrm{ad}^\mathrm{R} + \boldsymbol{\mathcal{A}}_\mathrm{ad}^\mathrm{xc},
\end{align}
where $\boldsymbol{\mathcal{A}}_\mathrm{ad}^\mathrm{R} = (\mathcal{U}^\dagger \boldsymbol{\mathcal{A}}^\mathrm{R} \,\mathcal{U})_\mathrm{ad}$.
Thus, the effective Hamiltonian for $\xc\to\infty$ contains the vector potential of a classical magnetic field which couples only to the orbital degree, accompanying the ``ferromagnetic" system~\cite{Bliokh2005,Gorini2010,Fujita2011}. It is given by the classical expression 
$
    \Beff = \nabla \times \boldsymbol{\mathcal{A}}_\mathrm{ad} = \Beff^\mathrm{R} + \Beff^\mathrm{xc}
$. By following this procedure, one finds the expressions
\begin{align}
  (\Beff^\mathrm{xc})_z & = -\frac{\hbar}{2	e}\sgn(\xc)~\hatn \cdot \left( \frac{\partial \hatn}{\partial x} \times \frac{\partial \hatn}{\partial y} \right) 
  \label{eq:tom_field}
 \\
  (\Beff^\mathrm{R})_z
 & =
 -\frac{ \effmass \soi}{e}\sgn(\xc)~\mathrm{div}~\hatn.
  \label{eq:com_field}
\end{align}
We thus arrive at the fundamental result that in addition to the field given by Eq.~(\ref{eq:tom_field}) above, which can be recognized as the generalization of Eq.~(\ref{eq:Beff}),
there is
a contribution to the overall field which explicitly depends on the chirality of the underlying texture and is non-vanishing already for one-dimensional spin textures.
In this context, it makes sense to refer to these co-existing 
fields as {\it topological} and {\it chiral} for $\Beff^\mathrm{xc}$ and $\Beff^\mathrm{R}$, respectively, see Fig.~(\ref{figure1:emergent_fields}).
Importantly, in contrast
to the emergent topological field, $(\beff^\mathrm{R})_z$, the local magnitude of $(\beff^\mathrm{R})_z$ is directly proportional to the strength of the spin-orbit interaction as given by $\soi$.
This appears to be very promising with respect to achieving a large magnitude of the chiral field in chiral spin textures emerging at surfaces and interfaces. To give a rough estimate, assuming a pitch of the texture on a length scale of
$L=\SI{20}{\nano\meter}$ and $\hbar\soi = \SI{1}{\electronvolt\angstrom}$ the amplitude of the local chiral emergent field reaches as much as
$2\pi\emass\soi/(eL)\approx \SI{270}{\tesla}$, which is roughly by an order of
magnitude larger than the corresponding topological field in a skyrmion of a similar size~\cite{Nagaosa2013}.

The emergence of two types of fields in spin textures, appearing in Eqs.~\ref{eq:tom_field} and \ref{eq:com_field}, is crucial for a qualitative understanding of the emergence of topological and chiral orbital magnetism, which are discussed in detail below. 

\begin{figure*}
 \centering
 \includegraphics[width=\linewidth]{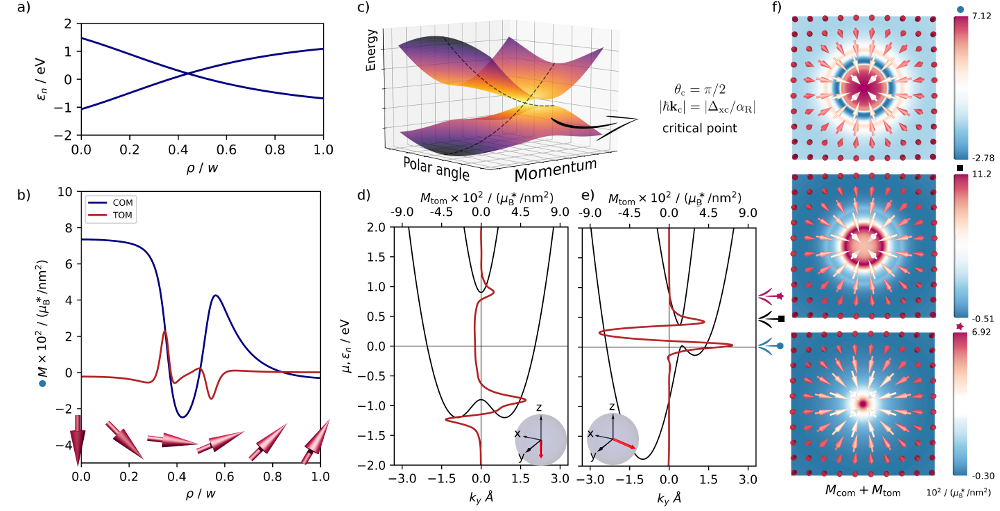}
 \caption{
 Interplay between real-space and $\vec{k}$-space topology for the case of a N\'eel skyrmion ($m=1$, $c=\SI{0}{\nano\meter}$, $w=\SI{20}{\nano\meter}$, $N_\mathrm{sk}=-1$). In all plots, we set $\hbar\soi = \SI{2}{\electronvolt\angstrom}$ and $\xc = \SI{0.9}{\electronvolt}$. b) $\com$ and $\tom$ show strong resonances which are correlated with a), a level crossing in the electronic structure when investigated as a function of $\rho$, the radial position inside the skyrmion. The nature of this crossing is further studied in the figures d) and e) which follow $\tom$ as a function of $\mu$ across the band-structure for two different positions in the skyrmion (indicated by the red arrow). The symbolic arrows on the right mark the values of $\mu$ used for the real-space distributions of $\com+\tom$ shown in f) and which exemplify the complex landscape of orbital magnetism in these systems for three different values of the chemical potential (indicated by the colored icons).
 }
 \label{fig:energy_dependence}
\end{figure*}


\vspace{0.4cm}
\noindent{\bf Chiral Orbital Magnetization}.
To get a first insight into the novel effect of COM, we consider the limit of small SOI, i.e., $\soi \ll \xc$. In this case, the gradient expansion (see Methods) provides an analytic expression for the local space-dependent orbital moment.
Up to $\mathcal{O}(\soi)$ it is given by
\begin{equation}
\label{com:model}
\com =  -\frac{1}{2} \chiLP  
 (\Beff^\mathrm{R})_z 
h(\mu/\xc)  , 
\end{equation}
where the function $h(x) \equiv  (3 x^2 - 1)  \Theta(1-|x|)/2$ describes the energy dependence of COM with $\Theta$ representing the Heaviside step-function. The magnitude of COM is thus directly proportional to the strength of spin-orbit interaction and vanishes in the limit of zero $\soi$. Furthermore, $\com$ is proportional to the diamagnetic Landau-Peierls susceptibility~\cite{Ashcroft1976} $\chiLP = -e^2 / (12\pi \effmass)$ which characterizes the orbital response of a free electron gas. Indeed, this seems reasonable in the limit of $\soi\ll\xc$ with the chemical potential positioned in the majority band, as the true orbital magnetic susceptibility of the Rashba model (as calculated by Fukuyama's formula~\cite{Fukuyama1971,OgataMasao2015}) reduces to 
$
\chi_\mathrm{oms}^{\phantom{\uparrow}} \sim \chiLP /2
$
in the same limit. For $| \mu | \approx | \xc| $ we therefore arrive at at the intuitive result guided by Eq.~(\ref{eq:tom_hypothesis}) with $\Beff$ replaced by the chiral emergent field:
\begin{equation}
\com =  -\frac{1}{2} \chi_{\mathrm{LP}}^{\uparrow + \downarrow } (\Beff^\mathrm{R})_z .
\label{com:emergent_picture}
\end{equation}
This reflects the fact that in the limit of $|\soi| \ll |\xc|$ the emergence of chiral orbital magnetization can be understood as the coupling of a mixed $SU(2)$ gauge field to the diamagnetic Landau-Peierls susceptibility.

The behavior of COM becomes complicated and deviates remarkably from the $\soi$-linear expression given by Eq.~(\ref{com:emergent_picture}) as the Rashba parameter increases. To demonstrate this, we numerically calculate the value of $\com$ at the center of a skyrmion, in a wide range of parameters $\xc$ and $\soi$ of the Rashba Hamiltonian, Eq.~(\ref{rashba:ham}), while fixing the chemical potential at $\mu=0$.
We parameterize the skyrmion in the polar coordinates $(\rho,\phi)$ by choosing
$ \hatn(\rho,\phi) = ( \sin\theta(\rho) \cos\Phi(\phi) , \sin\theta(\rho) \sin\Phi(\phi), \cos \theta(\rho))^T$ as the local magnetization vector\cite{Nagaosa2013}. Here, we define $\Phi(\phi) = m \phi + \gamma$ with the \emph{vorticity} $m$ and the \emph{helicity} $\gamma$. For a N\'eel skyrmion $\gamma =0 $, whereas a Bloch skyrmion is represented by the value $\gamma = \pi /2$. The topological charge of the skyrmion then equals $N_\mathrm{sk} = \int \dd x \dd y~\hatn\cdot( \partial_x \hatn \times \partial_y \hatn )/ (4 \pi) =-m$. In order to model the radial dependency refer to \citeauthor{Romming2015}\cite{Romming2015} and choose a $360^\circ$ domain wall, which is described by two parameters: the domain wall width $w$ and the core size $c$ (see Methods). 

The results are presented in Fig.~\ref{fig:phasediag} for a N\'eel skyrmion ($\gamma=0$) with $w = \SI{20}{\nm}$, $c = \SI{0}{\nm}$ and $m=1$. The magnetization is given  in units of $\effmagbohr / \si{\nano\meter}^2$ with the effective Bohr magneton $\effmagbohr = e \hbar / (2 \effmass)$. In this plot, we observe that while the gauge field picture is valid in the limit of $\xc / \soi \to \infty$, 
there exists a pronounced region in the ($\soi$,$\xc$)-phase-space
where COM exhibits a strong non-linear enhancement. This is in contrast to the case of Bloch skyrmions, where COM vanishes identically for all $(\soi,\xc)$, reflecting the symmetry of the Rashba coupling. It also elucidates our terminology, since already the gauge field description can be used to verify that $ \Com \propto \cos \gamma $ (for vorticity $m=1$), thereby making COM explicitly dependent on the helicity.


\vspace{0.4cm}
\noindent{\bf Topological Orbital Magnetization}.
The TOM appears as the correction to the OM which is second order in the gradients of the texture, Eq.~(\ref{eq:tom}), and while it vanishes for one-dimensional spin-textures, we show that it is finite for 2D textures such as magnetic skyrmions. 
In contrast to COM, the TOM is non-vanishing even without spin-orbit interaction. To investigate this, we set $\soi$ to zero, reducing the effective vector potential to $\boldsymbol{\mathcal{A}} = \boldsymbol{\mathcal{A}}^\mathrm{ xc}$ and with the emergent field turning into $ (\Beff^\mathrm{xc})_z$, Eq.~(\ref{eq:tom_field}). The gradient expansion (see Methods) now indeed reveals that
\begin{equation}
\tom  = -
\frac{1}{2} \chi_{\mathrm{LP}}^{\uparrow + \downarrow } (\Beff^\mathrm{xc})_z \
h(\mu/\xc),
\label{tom:emergent_picture}
\end{equation}
which again confirms the gauge-theoretical expectation. 
Remarkably, the similarity between Eqs.~(\ref{com:emergent_picture}) and (\ref{tom:emergent_picture}) underlines the common origin of the COM and TOM in the ``effective" magnetic field in the system, generated by a combination of a gradient of $\hatn$
along $x$ with spin-orbit interaction (in case of COM), and by a combination of the gradients
of $\hatn$ along $x$ and $y$ (in case of TOM).

{
\begin{figure}
\centering
\includegraphics[width=0.9\linewidth]{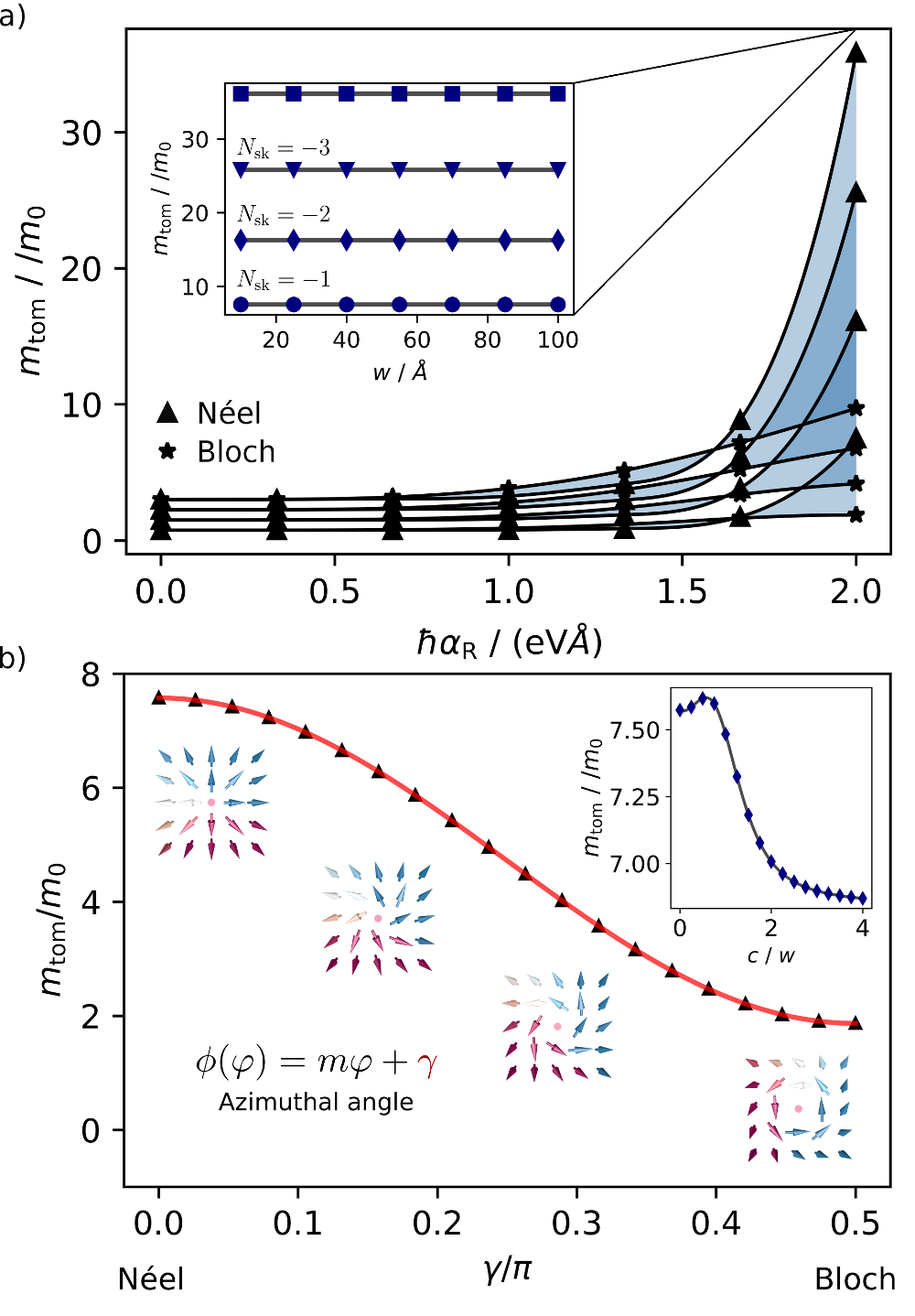}
\caption{a) When $\soi \ll \xc$, the integrated $\itom$ is a topological quantity which, in particular, cannot distinguish between topologically equivalent structures such as N\'eel (black triangles) and Bloch skyrmions (black stars).
Tracing the vorticities $m = 1,2,3,4$ as a function of $\soi$ and with $\xc = 0.9$, $\mu = 0$, the top figure illustrates how $\itom$ (in units of $m_0 = -\mu_\mathrm{B}^\ast/12$) passes from its regime of topological quantization ($\itom / m_0 = - m$) to a regime of strong enhancement, with N\'eel and Bloch structures clearly separated. Intermediate phases form a continuum between the two values (shaded regions). 
The inset demonstrates for the case of N\'eel skyrmions, that a level structure is still present at $\soi = \SI{2}{\electronvolt}$.
b) For the particular case of $\soi = 2.0$ and $\xc = 0.9,\mu = 0$ the $\gamma$ phase shift is used to interpolate from a N\'eel to a Bloch type Skyrmion, leading to a drastic loss of $\itom$. Variations in the shape of the skyrmion, as quantified by the ratio of $c/w$, have a very small effect.}
\label{fig:stability}
\end{figure}
}

To explore the behavior of TOM in the presence of spin-orbit interaction, $\soi \neq 0$, we numerically compute the value of TOM at the center of the N\'eel (Bloch) skyrmion with parameters used in the previous section, as function of $\xc$ and $\soi$ (at $\mu=0$). The corresponding phase diagram, presented in Fig.~\ref{fig:phasediag}, displays two notable features. The first one is the relative stability of Eq.~(\ref{tom:emergent_picture}) against a perturbation by a spin-orbit field in the limit of $|\xc| \gg |\soi|$. The second one is the significant enhancement of TOM in the regime where $|\soi| >|\xc|$,  similar to COM (albeit over a larger part of the parameter space). 


\vspace{0.4cm}

\noindent{\bf Interplay of topologies}. The phase diagrams in Fig.~\ref{fig:phasediag} have been evaluated at the core of a skyrmion. We now take a more global perspective and analyze the decomposition of the overall orbital magnetization into its constituent parts $\com$ and $\tom$ as a function of $\rho$, the radial position inside the skyrmion with $w = \SI{20}{\nano\meter}$ and $c=0$ (see Fig.~\ref{fig:energy_dependence}b)). By fixing $\hbar\soi$ to  $\SI{2}{\electronvolt\angstrom}$, and $\xc$ to $\SI{0.9}{\electronvolt}$ with $\mu =0$, we position ourselves precisely in the region of orbital enhancement discussed above in the context of the phase diagrams. 
When the local direction of the magnetization (parametrized by spherical coordinates $\theta$ and $\phi$) is close to the $z$-axis, $\com$ and $\tom$ are rather gently varying whereas their behaviour reveals strong resonance in the vicinity of in-plane directions ($\theta \approx \pi /2$). The emergence of this resonance coincides with an occurrence of a  band crossing at the critical $\vec{k}$-value of $\hbar k_\mathrm{c} = | \xc / \soi |$ with the polar coordinate in the Brillouin zone of  $\phi_\vec{k} = \phi - \pi /2$ in the local ferromagnetic electronic structure which corresponds to the given magnetization direction, see Fig.~\ref{fig:energy_dependence}a).

It is known that this specific band crossing in the Rashba model leads to a vastly enhanced diamagnetic susceptibility~\cite{Schober2012} and in close analogy, a strong response in $\com$ and $\tom$ can be expected based on Eq.~\ref{eq:tom_hypothesis}. To study the origin of this effect in greater detail, we plot $\tom$ as a function of $\mu$ for two different magnetization directions. The results, presented in Fig.~\ref{fig:energy_dependence}d) and e) reveal the sensitivity of $\tom$ to the SOI-mediated deformation of the purely parabolic free-electron bands separated by $\xc$. The magnitude of TOM is largest and exhibits pronounced oscillations in a narrow energy interval around the band edges. When we turn $\hatn$ into the in-plane direction, it can be seen how the resonances of $\tom$ are enhanced in magnitude and are carried along by those
band extrema which eventually touch at $\theta = \pi /2$, pushing the peaks of $\tom$ through the chemical potential (which was aligned to $\mu=0$ for Fig.~\ref{fig:energy_dependence}b)). For three different values of the chemical potential (indicated by the symbolic arrows) the strongly $\mu$-dependent real-space density of the total orbital magnetization $M$ is shown in  Fig.~\ref{fig:energy_dependence}f). This anisotropic behaviour cannot be accounted for within the emergent magnetic field picture which only relies on the real-space texture with its associated topological charge and winding density. 

The ``critical" metallic point in the Rashba model that we encounter is topologically non-trivial. Indeed, the upper and lower bands of the magnetic Rashba model bare non-zero Chern numbers,  $\mathcal{C}_1 = \pm \sgn(\xc) 1/2$, with the sign depending on the band~\cite{Shen2004}. The Chern number is a topological invariant of energy bands in $\vec{k}$-space and can only change when bands are crossing. 
Since the sign of $\mathcal{C}_1$ changes under the transformation $\xc \to -\xc$, the emergence of the critical metallic point at $\theta=\pi/2$ is enforced
when the direction of the magnetization is changed from $\theta = 0$ to $\theta =\pi$. This is illustrated in Fig.~\ref{fig:energy_dependence}c).
In the context of topological metals, such a point is known as a mixed Weyl point~\cite{Hanke2017a}, owing to the quantized 
flux of the Berry curvature permeating through the mixed space of $\vec{k}$ and $\theta$  (as confirmed explicitly by the calculations for the magnetic Rashba model). These points have recently been shown to give rise to an enhancement of spin-orbit torques and Dzyaloshinskii-Moriya interaction in ferromagnets~\cite{Hanke2017a}. Here, we demonstrate the crucial role that such topological features in the electronic structure could play for the pronounced chirality-driven orbital magnetism of spin textures.   
Given the observation that TOM simply follows the evolution of the electronic structure in real space via the direction of the local magnetization   
(as illustrated in a schematic way in Fig.~\ref{fig:energy_dependence}d)), 
the close correlation of real- and reciprocal space topologies offer promising design opportunities in skyrmions or domain walls of transition-metals with complex anisotropic electronic structure.


\vspace{0.4cm}
\noindent{\bf Topological Quantization and Stability}. One of the key properties of $\tom$ is its origin in the local real-space geometry of the texture. This has drastic consequences for the topological properties of the overall orbital moment of two-dimensional topologically non-trivial spin textures as we reveal below for the case of chiral magnetic skyrmions. We thus turn to the discussion of the total integrated values of the orbital moments in chiral spin textures by defining them as 
\begin{equation}
m_{\mathrm{com/tom}} = \int\dd\vec{x}~M_{\mathrm{com/tom}} (\vec{x}) . 
\end{equation}
Concerning the total value of the 
COM-driven orbital moment in one-dimensional uniform 360$^\circ$ or 180$^\circ$ chiral domain walls 
it always vanishes identically by arguments of 
symmetry (although it can be finite for example in a 90$^\circ$ wall). In sharp contrast, the TOM-driven total orbital moment of isolated skyrmions generally does not vanish. This can be most clearly shown in the limit when the gauge-field approach is valid (i.e., ~$\xc  \gg \soi$). In this case, the total flux of the emergent topological and chiral fields through an isolated skyrmion is given by
\begin{align}
  \Phi^\mathrm{xc} &\equiv \int_{\mathbb{R}^2}\dd\vec{x}~(\Beff^\mathrm{xc})_z = 2 \Phi_0 N_\mathrm{sk}
 \\
  \Phi^\mathrm{R} & \equiv \int_{\mathbb{R}^2}\dd\vec{x}~(\Beff^\mathrm{R})_z
  = 0.
  \label{eq:chiral_flux}
\end{align}
It then follows from Eq.~(\ref{com:emergent_picture}), 
that the integrated value of $\com$ vanishes while Eq.~(\ref{tom:emergent_picture}) predicts the {\it quantization} of the topological orbital moment $m_\mathrm{tom}$ to integer multiples of $\chiLP \Phi_0 = -\mu_\mathrm{B}^\ast / 6 $ (at $|\mu| = |\xc|$). 
In this limit, the skyrmion of non-zero topological charge $N_\mathrm{sk}\neq 0$ thus behaves as an ensemble of $N_\mathrm{sk}$ effective particles which occupy a macroscopic atomic orbital with an associated value of the orbital angular momentum of $\mu_\mathrm{B}^\ast/6$. This quantization is explicitly confirmed in Fig.~\ref{fig:stability}a), where we present the calculations of $m_\mathrm{tom}$ for N\'eel and Bloch-type skyrmions with dimensions $c=\SI{0}{\nano\meter} $ and $w=\SI{20}{\nano\meter}$ at a fixed value of $\xc=\SI{0.9}{\electronvolt}$ while varying $\soi$ for different topological charges
$N_\mathrm{sk} \in \lbrace -1,-2,-3,-4 \rbrace$. The results,  presented in units of $m_0 = -\mu_\mathrm{B}^\ast/12$ (corresponding to the Landau-Peierls limit at $\mu = 0$ and $N_\mathrm{sk} = -1$), reveal a stable plateau, corresponding to the regime of topological quantization, where $m_\mathrm{tom}$ attains the value $N_\mathrm{sk} m_0$. 

In the opposite limit of $\soi > \xc$ the magnitude of $\tom$ can
be enhanced drastically with respect to the topologically quantized value. When $\hbar\soi$ reaches a magnitude of about $\SI{1}{\electronvolt\angstrom}$, the emergent field picture breaks down and we discover a drastic increase in N\'eel--$m_\mathrm{tom}$ by as much as one order of magnitude upon increasing $\soi$. And although $m_\mathrm{tom}$ is not topologically quantized in this regime, it still attains a distinctly different magnitude for different skyrmion charges, and 
while it is weakly dependent on the $c/w$ ratio (up to a couple of percent), when $c=\mathrm{const}$ the variations of $w$ keep $m_\mathrm{tom}$ strictly constant (see the insets in Fig.~\ref{fig:stability} with $c=0$). The latter robustness can be demonstrated already analytically on the level of Eq.~(\ref{eq:tom}) using the transformation of coordinates $x \to x/w$.
Remarkably, in the regime of enhanced SOI, the strong dependence of the local TOM on the helicity of the skyrmion (i.e. N\'eel or Bloch), uncovered in Fig.~\ref{fig:phasediag}, is translated into a drastic dependence of the overall topological orbital moment on the way that the magnetization rotates from the core towards the outside region, as shown in Fig.~\ref{fig:stability}. Such behavior of the topological orbital moment with respect to deformations of the underlying texture suggests that monitoring the dynamics of the orbital magnetization in skyrmionic systems can be used not only to detect the formation of skyrmions with different charge, but also to distinguish various types of dynamical ``breathing" modes of skyrmion dynamics~\cite{Kim2014}.   
 

\vspace{0.4cm}
\noindent{\bf Discussion}\\
On a fundamental level, COM and TOM arise as a consequence of the changes in the local electronic structure caused by a non-collinear magnetization texture. Since the effective magnetic fields directly couple to the orbital degree of freedom, they lead to the emergence of chiral and topological orbital magnetization. While this intuitive interpretation in terms of real-space gauge fields eventually breaks down at large SOI, it makes room for a regime of strong enhancement in which the intertwined topologies of real- and reciprocal space lead to novel design aspects in the bandstructure engineering of orbital physics. This is possible by exploiting either the spin-orbit interaction or the dispersive behaviour of the bands, i.e. their effective mass. In particular, the metallic point in the mixed parameter space of the non-collinear Rashba model reveals its strong impact on COM and TOM. Such critical points will have a pronounced effect on the orbital magnetism even if they are emerging on the background of metallic bands in transition-metal systems. Our analysis therefore indicates in which materials an experimental detection of orbital magnetism associated with the non-collinearity is the most feasible. By numerically evaluating the magnitude and real-space behavior of the TOM and COM, we thereby  uncover that by tuning the parameters of surface and interfacial systems the orbital magnetism of domain walls and chiral skyrmions can be engineered in a desired way. 


Concerning experimental observation of the effects discussed here, $\tom$ and $\com$ could be accessible by techniques such as off-axis electron holography~\cite{Shibata2017} (sensitive to local distribution of magnetic moments), or scanning tunneling spectroscopy (sensitive to the local electronic structure) in terms of $B$-field induced changes in the $\dd I/\dd U$ or $\dd^2I/\dd U^2$ spectra~\cite{Kubetzka2017}. 
An already existing proposal for the detection of non-collinearity driven orbital magnetization of skyrmions by \citeauthor{Dias2016} relies on X-ray magnetic circular dichroism (XMCD) which is able to distinguish orbital contributions to the magnetization from the spin contributions\cite{Dias2016}.

Further, the emergence of COM and TOM can give a thrust to the field of electron vortex beam microscopy~\cite{Fujita2017} -- where a beam of incident electrons intrinsically carries orbital angular momentum interacting with the magnetic system -- into the realm of chiral magnetic systems. For example, we speculate that at sufficient intensities, electron vortex beams could imprint skyrmionic textures possibly by partially transforming its orbital angular momentum into TOM. Since the topological orbital moment is directly proportional to the topological charge of the skyrmions, we also suggest that the interaction of TOM with external magnetic fields could be used to trigger the formation of skyrmions with large topological charge. Ultimately, the currents of skyrmions can be employed for low-dissipation transport of the associated topological orbital momenta over large distances in skyrmionic devices. 


While in this work we focus primarily on TOM, the here discovered  chiral orbital magnetization has been an overlooked quantity in chiral magnetism so far. 
Besides the fact that it emerges already in one-dimensional chiral systems
and serves as a playground to study the effects of mixed space Berry phases, it can reach very large values depending on
the details of the texture as well as strength of SOI.
Even in case of skyrmions, where the argument of vanishing effective flux, Eq.~\ref{eq:chiral_flux}, might suggests that COM is not of importance, it turns out that beyond the $\soi\to0$ limit the integral effect of $\com$ can be substantially enhanced in a way similar to TOM. 
A prominent example for the importance of COM is given~e.g.~in Vanadium-doped BiTeI~\cite{Klimovskikh2017} which has a large SOC of $\hbar\soi = \SI{3.8}{\electronvolt\angstrom}$ and an exchange gap of $\xc = \SI{45}{\milli\electronvolt}$. If this material would host N\'eel skyrmions ($m=1$, $w= \SI{20}{\nano\meter}$, $c = \SI{0}{\nano\meter}$), $\icom$ would reach approximately $12 \effmagbohr$ which is magnitudes larger than the corresponding $\itom$ of about $-0.7 \effmagbohr$. Creating skyrmions and large COM in strong Rashba systems might therefore be a promising direction to pursue.


In a wider perspective, the emergence of TOM and COM gives rise to a physical object which is directly connected to the orbital degree of freedom with the advantage that it can be understood from a semiclassical perspective in a way which is engineerable and controllable. Our findings thus open new vistas for exploiting the orbital magnetism in chiral magnetic systems, thereby opening interesting prospects for the field of ``chiral" spintronics and  orbitronics.

\vspace{0.5cm}
\noindent{\bf Methods}
{
\small
\\
{\bf Gradient expansion}. The expansion in exchange field gradients is naturally achieved within the phase-space formulation of quantum mechanics, the Wigner representation~\cite{Onoda2006, Freimuth2013}. The key quantity in this approach is the retarded single-particle Green's function $\Gret$, implicitly given by the Hamiltonian $H$ via the Dyson equation
\begin{equation}
\left(
\epsilon
- H(\vec{x},\boldsymbol{\pi} ) + i0^+
\right)\star \Gret(x,\pi) = \id ,
\label{eq:dyson}
\end{equation}
where $x^\mu = (t,\vec{x})$ and $\pi^\mu = (\epsilon,\boldsymbol{\pi})$ are the four-vectors of position and canonical momentum, respectively. The latter of the two, in terms of the elementary charge $e>0$ and the electromagnetic vector potential $A$, is related to the zero-field momentum $p$ by the relation 
$
\pi_\mu (x,p) = p_\mu + e A_\mu(x) .
$
The $\star$-product, formally defined by the operator
\begin{equation}
\star \equiv \exp \left\lbrace\frac{i \hbar}{2}  \left(
\overset{\leftarrow}{\partial}_{x^\mu}   \overset{\rightarrow}{\partial}_{\pi_\mu} 
-
\overset{\leftarrow}{\partial}_{\pi_\mu}   \overset{\rightarrow}{\partial}_{x^\mu} 
- e F^{\mu\nu}\overset{\leftarrow}{\partial}_{\pi^\mu}
\overset{\rightarrow}{\partial}_{\pi^\nu}
 \right)   \right\rbrace
\end{equation}
of left- and right-acting derivatives $\tiny\overset{\leftrightarrow}{\partial}$, allows for an expansion of $\Gret$ in powers of $\hbar$, gradients of $\hatn$ and external electromagnetic fields, captured in a covariant way by the  field tensor $F^{\mu\nu} = \partial_{x_\mu} A^\nu - \partial_{x_\nu} A^\mu$~\cite{Onoda2006}.

In this work, we are after the orbital magnetization (OM) in $z$-direction. Given the grand canonical potential $\Omega$, the surface density of the orbital moment is given by~\cite{Zhu2012,Shi2007}
\begin{equation}
M(x) = - \partial_B \Braket{\Omega(x)} ,
\label{eq:magformula}
\end{equation}
which requires an expansion of $\Omega$ up to at least first order in the magnetic field $\mathbf{B} = B \vec{e}_z$ in the collinear case. In the limit of $T\to 0 $, the grand potential is asymptotically related to the Green's function $\Gret$ via
\begin{equation}
 \Braket{\Omega} \sim - \frac{1}{\pi}
 \ \Im\   \int \frac{\dd p} {(2 \pi \hbar)^2}   \ f(\epsilon)  (\epsilon - \mu) \ \tr\ \Gret (x,p),
 \label{eq:grandpot}
\end{equation}
where $\Im$ denotes the imaginary part, the  integral measure is defined as  $\dd p = \dd \epsilon\ \dd^2\mathbf{p}$, $f(\epsilon)$ represents the Fermi function $f(\epsilon) = (e^{\beta (\epsilon-\mu) } +1 )^{-1}$, $\mu$ is the  chemical potential and $\beta^{-1} =k_\mathrm{B} T $. In our approach, deviations from the collinear theory enter the formalism as gradients of $\hatn$ and can be traced systematically in $\Gret$ and in $\Braket{\Omega}$, finally leading to Eq.~(\ref{eq:com}) and (\ref{eq:tom}).

\vspace{0.4cm}
\noindent{\bf Computational details}. All calculations were performed with a Green's function broadening $i0^+ \to i \Gamma$ with $\Gamma = \SI{100}{\milli\electronvolt}$ while we approach the zero-temperature limit by setting $ k_\mathrm{B} T = \SI{10}{\milli\electronvolt}$. The $\vec{k}$-space integrals are then performed on a quadratic $512\times512$ mesh. The effective electron mass was set to $\effmass/ \emass = 3.81$ everywhere except for the example of V-doped BiTeI, where $\effmass/ \emass = 0.1$~\cite{Ishizaka2011}.

\vspace{0.4cm}
\noindent{\bf Skyrmion parametrization}. In order to model the skyrmions in this work we choose the profile which can be described by the parametrization\cite{Nagaosa2013} 
\begin{equation}
\hat{\mathbf{n}}(\rho,\phi) = \begin{pmatrix}
 \sin(\theta(\rho)) \cos(\Phi(\phi)) \\
  \sin(\theta(\rho)) \sin(\Phi(\phi))\\
  \cos(\theta(\rho)) 
\end{pmatrix}.
\end{equation}
The topological charge is then given by
\begin{align}
N_\mathrm{sk} &= \frac{1}{4 \pi}
\int \dd x \int \dd y\  \hat{\mathbf{n}} \cdot \left( \partial_x \hat{\mathbf{n}} \times \partial_y \hat{\mathbf{n}} \right) 
\notag \\
&=
 -\frac{1}{4 \pi}
\Phi(\phi) \Big|_{0}^{2\pi}  \cos \theta(\rho) \Big|_{0}^{\infty}. 
\end{align}
Assuming $\Phi(\phi)  = m \phi + \gamma$, with the vorticity  $m\in\mathbb{Z}$ and the helicity $\gamma\in\mathbb{R}$ (N\'eel skyrmions correspond to $\gamma=0$, Bloch skyrmions to $\gamma = \pi/2$), as well as the property $\theta(0)=\pi$ and $\theta(\infty) = 0$, the integral evaluates to $N_\mathrm{sk} = - m $. A realistic profile satisfying these requirements and which is used in this work is given by~\cite{Romming2015}
\begin{align}
\theta(\rho) & = \sum_\pm \arcsin \left( \tanh \left( - \frac{-\rho \pm c}{w / 2} \right) \right) + \pi,
\end{align}
with the core size $c$ and the domain wall width $w$. 

\vspace{0.4cm}
\noindent{\bf Data availability}. The code and the data that support the findings of this work are available from the corresponding authors on request.
}

\vspace{0.5cm}
\def\bibsection{\noindent{\bf References}}

\bibliography{letter}
\bibliographystyle{apsrev4-1}

\vspace{0.5cm}
\noindent{\bf Acknowledgements}
\\
We thank J.-P. Hanke, M.d.S. Dias and S. Lounis for fruitful discussions, and gratefully acknowledge computing time on the supercomputers JUQUEEN and JURECA at Jülich Supercomputing Center, and at the JARA-HPC cluster of RWTH Aachen. We acknowledge funding under SPP 1538 and project MO 1731/5-1 of Deutsche Forschungsgemeinschaft (DFG) and the European Union's Horizon 2020 research and innovation programme under grant agreement number 665095 (FET-Open project MAGicSky). This work has been also supported by the DFG through the Collaborative Research Center SFB 1238 and the priority programme SPP 2137.

\newpage
\vspace{0.5cm}
\noindent{\bf Author contributions}
\\
F.R.L. uncovered the emergence of chiral and topological orbital magnetization from the semiclassical expansion. F.R.L. and Y.M. wrote the manuscript. All authors discussed the results and reviewed the manuscript.

\vspace{0.5cm}
\noindent{\bf Additional information}
\\
{\bf Competing financial interests.} The authors declare no competing financial interests.


\end{document}